\documentclass[aps,preprint,amsmath,amssymb]{revtex4}
\usepackage{graphicx,color}
\usepackage{subfigure}
\def\be{\begin{eqnarray}}
\def\ed{\end{eqnarray}}
\def\non{\nonumber}

\def\lam{\lambda}

\begin{document}


\title{ Two-Higgs-Doublet  Type-II Seesaw Model}

\author{ \bf Chuan-Hung Chen$^{a}$\footnote{Email:
physchen@mail.ncku.edu.tw} and Takaaki Nomura$^{a}$\footnote{Email: nomura@mail.ncku.edu.tw} }

\affiliation{ $^{a}$Department of Physics, National Cheng-Kung
University, Tainan 701, Taiwan  }

\date{\today}

\begin{abstract}

 Motivated by the new observed scalar boson of 126 GeV at  ATLAS and CMS, various phenomena  in two-Higgs-doublet model (THDM)  are investigated broadly in the literature. For considering the model that possesses a solution to the massive neutrinos, we study the simplest extension of conventional type-II seesaw model to two Higgs doublets. We find that the new interactions in the scalar potential  cause the  sizable mixture of charged Higgses in triplet and doublet. As a result, we have a completely different decay pattern for doubly charged Higgs ($\delta^{\pm\pm}$), even the vacuum expectation value (VEV) of Higgs triplet is at  GeV level, which is limited by the precision measurement for  $\rho$-parameter. For illustrating the new characters of the model, we study the influence of new interactions on the new open channels $\delta^{++}\to ( H^+_1 W^{+^{(*)}}, H^+_1 H^+_1)$ with $H^+_1$ being the lightest charged Higgs. Additionally, due to the new mixing effect, the triplet charged Higgs could couple to quarks in the model; therefore, the search for $\delta^{++}$  via $\delta^{++}\to tb W^+ \to b \bar b W^+ W^+$ by mediated $H^+_{1}$ becomes significant.

\end{abstract}

\maketitle

The recent observation of a new scalar particle at 126 GeV by ATLAS~\cite{:2012gk} and CMS~\cite{:2012gu}  shows  that  the Higgs mechanism is  a right direction not only for the origin of masses of gauge bosons  but also for the masses of quarks and charged leptons in the standard model (SM). By this point of view, the most mysterious observed phenomenon in  particle physics is the masses of neutrinos. Besides the undetermined mechanism of neutrino masses, we also know nothing about their mass ordering, which is classified by normal ordering, inverted ordering, and  quasi-degeneracy in the literature~\cite{PDG2012}.

Before the observations of neutrino oscillations, numerous mechanisms for generating the neutrino masses had been proposed.  For instance,  type-I seesaw \cite{SeeSaw} mechanism  introduced the heavy right-handed neutrinos while the type-II seesaw mechanism \cite{ Magg:1980ut,Konetschny:1977bn}  extended the SM by including a $SU(2)$ Higgs triplet.  Additionally, other possibilities were also investigated such as  adding the new  triplet fermions~\cite{Foot:1988aq}, radiative corrections~\cite{Ma:2006km,Krauss:2002px,Aoki:2008av}, etc. Due to the similarity in mass generation mechanism between type-II seesaw  and Higgs mechanism, we focus the  study on the simplest extension to the type-II seesaw model.

The characters of type-II seesaw model with one Higgs doublet and one Higgs triplet can be briefly summarized as follows: first, doubly charged Higgs decays to the same sign charged gauge bosons (WW) and  leptons ($\ell\ell$), where the former coupling is associated with vacuum expectation value (VEV) of triplet denoted by $v_\Delta$ and the latter is related to the multiplication of Yukawa couplings and $v_\Delta$. The involved parameters are limited to be small by the observed neutrino masses.   
 Second, for achieving the small $v_\Delta$, one needs to require either a small massive coupling for  $H^T i \tau_2 \Delta^\dagger H$ term or a heavy mass scale for Higgs triplet;  we will see this point later.  
If we adopt the mass scale of Higgs triplet  to be of $O(100)$ GeV, 
it is then inevitable to have a hierarchy in the massive parameters of Lagrangian. 
%
For instance, 
%
 if one requires  leptonic decays of doubly charged Higgs to be dominant, because of the requirement of vacuum stability, the coefficient $\mu$ of  $H^T i\tau_2 \Delta^\dagger H$  term  in the scalar potential has to be $\mu\sim v_\Delta < 10^{-4}$ GeV. 
%
%
Third,  the singly charged Higgs of triplet does not couple to  quarks.
 

 From theoretical viewpoint, the two-Higgs-doublet model (THDM) was proposed  for solving the weak and strong CP problems~\cite{Lee:1973iz,Peccei:1977hh}. In spite of the original motivation, THDM itself provides rich phenomena in particle physics. By the new discovery of  126 GeV scalar boson at ATLAS and CMS, the phenomenology of THDM has been further investigated broadly in the literature, e.g. Refs.~\cite{ Ferreira:2012nv,Diaz-Cruz:2014aga,Barger:2014qva}. Since the THDM does not have the mechanism to generate the masses of neutrinos, according to the discussions on the conventional type-II seesaw model (CTTSM), the massive neutrinos indeed could originate from a Higgs triplet with a non-vanished  VEV.  
Therefore, in this paper, we study the extension of CTTSM by  including one extra Higgs doublet, i.e. two-Higgs-doublet (THD) and one Higgs triplet model.  We find that unlike the case in CTTSM, the couplings $\mu_j $ of  $H^T_{j} i \tau_2 \Delta^\dagger H_k $ terms in the scalar potential could be as large as electroweak scale when the small $v_\Delta$ is satisfied. Moreover, the $\mu_j$ terms cause
the new mixing effects in singly charged Higgses and new decay channels for doubly and singly charged Higgses.
 Then the mixing of charged Higgses from doublets and triplet can be large, which is small in CTTSM due to the small $\mu$ coupling.
Consequently, these new effects will change the search of doubly charged Higgs at colliders~\cite{Han:2007bk, Akeroyd:2007zv, delAguila:2008cj, Aoki:2011pz, Chiang:2012dk,Sugiyama:2012yw,Kanemura:2013vxa,Chun:2013vma, Melfo:2011nx,Chun:2003ej,Akeroyd:2005gt,Perez:2008ha,Arhrib:2011uy} and affect  the rare decays in low energy physics, such as $b\to s \gamma$, $B\to \tau \nu$, $B\to D^{(*)} \tau \nu$, etc. \cite{Chen:2006nua}. 
 
In order to better understand the new characters of the extended model, in the following we briefly introduce the model. The involved Higgs doublets and triplet are denoted by $H_{1,2}$ and $\Delta$, respectively. Their representations in $SU(2)$ group are chosen as
 \be
H_1 &=&\left( \begin{array}{c}
   H^+_1 \\
   (v_1+ \rho_1 + i\eta_1 )/\sqrt{2} \\
     \end{array} \right)\,, ~~~ H_2=\left( \begin{array}{c}
   H^+_2 \\
   (v_2 + \rho_2+i\eta_2)/\sqrt{2} \\
     \end{array} \right)\,, \non \\
     \Delta &=& \left( \begin{array}{cc} 
    \delta^+/\sqrt{2} & \delta^{++}  \\ 
    (v_\Delta + \delta^0 + i\eta^0)/\sqrt{2} & -\delta^+/\sqrt{2} \\ 
 \end{array}  \right)\,, \label{eq:rep}
 \ed
 where $v_{1,2,\Delta}$ stand for the VEVs of neutral components of $H_1$, $H_2$ and $\Delta$ respectively.
 As known, the general THDM will cause flavor changing neutral currents (FCNCs) at tree level in Yukawa sector.  For avoiding the FCNC effects, we impose a $Z_2$ symmetry at the Yukawa interactions. Under the symmetry,  the transformations of matter fields are given by 
 \be
 H_2 \to - H_2\,, ~~~ U_R \to -U_R
 \ed
 with $U_R$ being the right-handed up-type quarks. The other fields are unchanged in the $Z_2$ transformation.   Accordingly, the Yukawa couplings are written by
 \be
 -{\cal L}_Y &=& \bar Q {\bf Y^d} D_R H_1 + \bar Q {\bf Y^u}  U_R \tilde H_2 + \bar L {\bf Y^\ell} \ell_R H_1 \non \\
 &+& \frac{1}{2}\left[ L^T C{\bf h}  i\sigma_2 \Delta P_L L +h.c. \right] \,, \label{eq:lang_y}
 \ed
where we have suppressed all flavor indices, $Q^T=(u, d)_L$ and $L^T=( \nu, \ell)_L$ are the $SU(2)_L$ doublets of quarks and leptons, $(D_R,U_R,\ell_R)$  in turn denotes  the $SU(2)_L$ singlet for down-type, up-type quarks and charged leptons, and $\tilde H=i\sigma_2 H^*$ with $\sigma_2$ being the second Pauli matrix. The detailed discussions for the Yukawa couplings could refer to Ref.~\cite{Chen:2006nua}. Since the signal of doubly charged Higgs is clearer and unique in type-II seesaw model,  in this study we will focus on the  decays associated with $\delta^{\pm\pm}$. By Eq.~(\ref{eq:lang_y}), the relevant interactions with leptons are given by
 \be
 {\cal L}_{\delta^{\pm\pm} \ell \ell} &=& \frac{1}{2} \ell^T C {\bf h} P_L \ell \delta^{++} +h.c. \,, \non \\
 {\bf h}&=& \frac{\sqrt{2}}{v_{\Delta}} U^*_{\rm PMNS} {\bf m}^{\rm dia}_{\nu}U^{\dagger}_{\rm PMNS} \,.
 \label{eq:int_y}
 \ed
 Here, ${\bf m}^{\rm dia}_\nu$ is the diagonalized neutrino mass matrix and $U_{\rm PMNS}$ is the Pontecorvo-Maki-Nakagawa-Sakata (PMNS) matrix \cite{Pontecorvo:1957cp,Maki:1962mu}. From Eq.~(\ref{eq:int_y}), one can see that the typical coupling of $\delta^{\pm\pm}$ to lepton-pair is proportional to $m_\nu /v_\Delta$. Consequently, if we take the masses of neutrinos as the knowns which are determined by experiments, the partial decay rate for $\delta^{\pm\pm}\to \ell^\pm \ell^\pm$ strongly depends on the value of $v_\Delta$.

Besides the leptonic couplings, $\delta^{\pm\pm}$ also couples to charged gauge boson and the couplings could be read from the gauge  invariant kinetic terms of Higgs fields. Hence, we write the kinetic terms as 
\begin{align}
{\cal L}_{\rm K.T.} = (D^\mu H_1)^\dagger (D_\mu H_1) + (D^\mu H_2)^\dagger (D_\mu H_2)+ Tr \left[ (D_{\mu} \Delta)^{\dagger} D^{\mu} \label{eq:lang_kt}\Delta \right].
\end{align}
The covariant derivatives of the associated fields are expressed by 
\begin{align}
D_\mu H_{1(2)} &= \left( \partial_\mu - i \frac{g}{\sqrt{2}}(W_\mu^+ T^+ + W_\mu^- T^-) -i \frac{g}{C_W} Z_\mu (T^3-S_W^2 Q) -ie A_\mu Q \right) H_{1(2)} \,, \non \\
D_{\mu} \Delta &= \partial_{\mu} \Delta -i \frac{g}{\sqrt{2}} \bigl( W_{\mu}^+ [T^+,\Delta]+W_{\mu}^-[T^-,\Delta] \bigr)\,, \non \\
&
-i \frac{g}{c_W} Z_{\mu} \bigl( [T^3,\Delta]-S_W^2 [Q,\Delta] \bigr) -ie A_{\mu} [Q,\Delta]\,, \label{eq:cov}
\end{align}
where the $W^\pm_\mu$, $Z_\mu$ and $A_\mu$ stand for the gauge bosons in the SM, $g$ is the gauge coupling constant of the SU(2), $e$ is the electromagnetic coupling constant, 
$S_W(C_W) = \sin \theta_W(\cos \theta_W)$ with $\theta_W$ being the Weinberg angle,
$T^\pm = (\sigma^1 \pm i \sigma^2)/2$ and $T^3 = \sigma^3/2$ are defined by the Pauli matrices $\sigma^i$, and $Q$ is the electric charge operator. After electroweak symmetry breaking (EWSB), the masses of $W^\pm$ and $Z$ bosons are obtained by
 \be
 m^2_W &=& \frac{g^2 v^2}{4} \left( 1 + \frac{2v^2_\Delta }{v^2} \right) \,, \non \\
 m^2_Z &=& \frac{g^2 v^2}{4\cos^2\theta_W} \left( 1 + \frac{4 v^2_\Delta}{v^2 }\right)
 \ed
 with $v=(v^2_1+v^2_2)^{1/2}$.
As a result, the $\rho$-parameter at tree level could be obtained as
 \be
 \rho= \frac{m^2_W}{m^2_Z c^2_W} = \frac{ 1+ 2 v^2_\Delta/ v^2} { 1+4 v^2_\Delta/v^2}\,.
 \ed
Taking the current precision measurement for $\rho$-parameter to be $\rho=1.0004^{+0.0003}_{-0.0004}$ \cite{PDG2012}, we get  $v_{\Delta} < 3.4$ GeV when 2$\sigma$ errors are taken into account.  By Eqs.~(\ref{eq:rep}), (\ref{eq:lang_kt}) and (\ref{eq:cov}), the interactions of $\delta^{\pm\pm}$ with $W^\mp$ are found by 
\be
{\cal L}_{\delta^{\pm\pm} W^{\mp} } =-ig (\partial_\mu \delta^{++}) \delta^{-} W^{-\mu} + i g \delta^{++} (\partial_\mu \delta^-) W^{-\mu} + \frac{1}{2} \left(\sqrt{2} g^2 v_\Delta\right)\delta^{++} W^-_\mu W^{-\mu} + h.c. \label{eq:d++W}
\ed
We see clearly that the coupling of $\delta^{++}$ to $W^-W^-$ is proportional to $v_\Delta$. In CTTSM, the value of $v_\Delta$ determines which decaying channel is the dominant mode, $\ell \ell$ or $WW$ channel.  Since $\delta^{\pm\pm}$ and $\delta^\pm$ belong to the same multiplet and get the masses from $m_\Delta$ before EWSB, the possible mass difference $m_{\delta^{\pm\pm}}- m_{\delta^\pm}$ is at most of $O(m_{W})$.  Therefore, for $m_{\delta^{++}} > m_{\delta^+}$, the decay $\delta^{++}\to W^+ \delta^+$ is suppressed by phase space. However, by the first two interactions  in Eq.~(\ref{eq:d++W}), where the couplings are independent of $v_\Delta$,  the three body decay $\delta^{++}\to \delta^{+} W^{+^*} ( \to \ell^{+} \nu)$ indeed can be significant~\cite{Melfo:2011nx,Chun:2003ej,Akeroyd:2005gt,Perez:2008ha}. Nonetheless, when a new charged Higgs is introduced, we will show that the new interactions  in scalar potential will lead to a different decay pattern for doubly charged Higgs . 

In the following we give detailed discussions on the scalar potential, which is the origin of the crucial effects in our model. The scalar potential of THD and triplet in $SU(2)_L\times U(1)_Y$ symmetry is expressed as
 \be
 V(H_1, H_2, \Delta) &=&  V_{H_1H_2} + V_\Delta + V_{H_1 H_2 \Delta}\,, \label{eq:v1}
 \ed
 where $V_{H_1 H_2}$ and  $V_\Delta$  stand for the scalar potential of THDM and of pure triplet, and  $V_{H_1 H_2 \Delta}$is the interaction among $H_1$, $H_2$ and $\Delta$.  Their expressions are given by 
  \be
V_{H_1H_2} &=& m^2_1 H^\dagger_1 H_1 + m^2_2 H^\dagger_2 H_2 - m^2_{12} ( H^\dagger_1 H_2 + h.c.)+ \lam_1 ( H^\dagger_1 H_1)^2  \non \\
 &+&  \lam_2 (H^\dagger_2 H_2)^2 + \lam_3  H^\dagger_1 H_1 H^\dagger_2 H_2+ \lam_4  H^\dagger_1 H_2 H^\dagger_2 H_1 + \frac{\lam_5}{2} \left[(H^\dagger_1 H_2)^2+h.c. \right] \,, \non \\
 V_\Delta &=& m^2_\Delta Tr \Delta^\dagger \Delta + \lam_{9} (Tr \Delta^\dagger \Delta)^2 + \lam_{10} Tr (\Delta^\dagger \Delta)^2\,, \non \\
 V_{H_1H_2\Delta} &=& \left( \mu_1 H^T_1 i\tau_2 \Delta^{\dagger}  H_1 + \mu_2 H^T_2 i \tau_2 \Delta^\dagger H_2 + \mu_3 H^T_1 i\tau_2 \Delta^\dagger H_2 + h.c. \right) \non \\
 &+& \left( \lam_6 H^\dagger_1 H_1 + \bar\lam_6 H^\dagger_2 H_2 \right) Tr \Delta^\dagger \Delta + H^\dagger_1 \left( \lam_7 \Delta \Delta^\dagger + \lam_8 \Delta^\dagger \Delta \right) H_1 \non \\
 &+& H^\dagger_2 \left( \bar\lam_7 \Delta \Delta^\dagger + \bar\lam_8 \Delta^\dagger \Delta \right) H_2\,. \label{eq:v2}
 \ed
We note that  the imposed $Z_2$ symmetry is broken spontaneously. In order to make the ultraviolte divergences of  higher order effects under control, as usual, we keep the $Z_2$ soft breaking terms $m^2_{12} H^\dagger_1 H_2$ and $\mu_3 H^T_1 i\sigma_2 \Delta^\dagger H_2$ in the scalar potential, in which  the former is mass dimension 2 while the later is mass dimension 3; however,   the $Z_2$ hard breaking terms are suppressed. Since we will not discuss the CP violating effects, hereafter, we take all couplings in the potential as real values.  By Eqs.~(\ref{eq:v1}) and (\ref{eq:v2}), the VEVs of neutral scalar fields could be determined  by the minimal conditions $\partial \langle V\rangle / \partial v_{1,2,\Delta}=0$. As a result, we have
 \be
 \frac{\partial \langle V \rangle}{\partial v_1} &\approx&  m^2_1 v_1 -m^2_{12} v_2 + \lam_1 v^3_1 + \lam_L v^2_2 v_1 \approx 0\,, \non \\
 \frac{\partial \langle V \rangle }{\partial v_2} &\approx& m^2_2 v_2 - m^2_{12} v_1 + \lam_2 v^3_2 + \lam_L v^2_1 v_2  \approx 0\,, \non \\
 \frac{\partial \langle V \rangle }{\partial v_\Delta} &\approx& m^2_\Delta v_\Delta  -\frac{1}{\sqrt{2}} \left( v^2_1 + \mu_1 + v^2_2 \mu_2 + v_1 v_2 \mu_3\right) \non \\ 
 &+& \left[ \frac{\lam_6 +\lam_7 }{2} v^2_1  
 + \frac{\bar \lam_6 + \bar\lam_7}{2} v^2_2 \right] v_\Delta \approx 0\,, \label{eq:mini}
 \ed
 where the terms associated with  $v_\Delta $ in the first two equations and  $ v^3_\Delta $ in the third equation have been ignored due to  $v_\Delta \ll v_{1, 2}$. From the last equation, the VEV of neutral triplet is obtained by
  \be
  v_\Delta \approx \frac{1}{\sqrt{2}} \frac{\mu_1 v^2_1 + \mu_2 v^2_2 + \mu_3 v_1 v_2}{m^2_\Delta + (\lam_6+\lam_7) v^2_1/2 + (\bar\lam_6+\bar\lam_7)v^2_2 /2}\,. \label{eq:v_d}
  \ed
By this result, we see that  with $\mu_2=\mu_3 =0$, the small $v_\Delta$ indicates  the small $\mu_1$ or large  $m_\Delta$ in CTTSM.  However, when the $\mu_2$ and $\mu_3$ effects are introduced, the necessity  of small $v_\Delta$ could be accommodated by  the massive parameters $\mu_{1,2,3}$ and $m_\Delta$, which can be in the same order of magnitude. Hence, the magnitude of $v_\Delta$ indeed could be adjusted by the free parameters of the new scalar potential 
without introducing a hierarchy to the massive parameters. 
%
  %

By counting the physical degrees of freedom, we have three CP-even neutral particles, two CP-odd pseudoscalar bosons, two singly charged Higgses and one doubly charged Higgs in the model.  The new interactions such as $\mu_{1,2,3}$ terms in $V_{H_1 H_2 \Delta}$
could cause interesting effects on the couplings of SM-like Higgs, pseudoscalars, charged Higgses, and doubly charged Higgs; moreover, their producing  and decaying channels are also modified. For illustrating the features of this model, we  concentrate on the new mixing effects of singly charged Higgses and on the new decaying channels of doubly charged Higgs. The complete analysis of the model will be given elsewhere. 

We have shown the couplings of $\delta^{\pm\pm}$ to leptons and W-gauge boson in Eqs.~(\ref{eq:lang_y}) and (\ref{eq:d++W}). For discussing the singly charged Higgs effects, like conventional THDM, we combine both doublets $H_1$ and $H_2$ to be
\be
\bar h&=&\cos\beta H_1 + \sin\beta H_2 =   \left(\begin{array}{c}
    G^+ \\ 
    (v+ h^0 + i G^0)/\sqrt{2} \\ 
  \end{array} \right)\,, \non \\
\bar H&=& -\sin\beta H_1 + \cos\beta H_2 = \left(\begin{array}{c}
    H^+ \\ 
    (H^0 + i A^0)/\sqrt{2} \\ 
  \end{array} \right)\,, \label{eq:hH}
\ed
where only the doublet $\bar h$ has the VEV after EWSB and $\sin\beta (\cos\beta)= v_2/v (v_1/v)$.  As known, in THDM $h^0$ and $H^0$ are the CP-even scalars and they are not physical states, $A^0$ is the physical  CP-odd scalar boson, and $H^\pm$ is the physical charged Higgs particle. When the $SU(2)$ triplet $\Delta$ is included to the model,  $\delta^0$, $\eta^0$ and $\delta^{\pm}$ of $SU(2)$ triplet will mix with $(h^0, H^0)$, $A^0$ and $H^\pm$, respectively. 
In this study, we will concentrate on the new  mixing effects of charged Higgses and their implications. 
 For simplifying numerical analysis and preserving the requirement of $v_\Delta << v_1, v_2$, we adopt the relation 
 \be
 \mu_3 \sim - \frac{\mu_1 v^2_1 + \mu_2 v^2_2  }{v_1 v_2}\,. \label{eq:mu3}
 \ed
For completeness,  we also show the mass matrices of CP-odd and CP-even Higgs bosons in the appendix. Hence, in terms of the triplet representation in Eq.~(\ref{eq:rep}), doublet representations in Eq.~(\ref{eq:hH}) and  scalar potentials in Eq.~(\ref{eq:v2}),  the  mass matrix for $G^+$, $H^+$ and $\delta^+$ is written by
 \be
  (G^- H^- \delta^- ) \left( \begin{array}{ccc}
    0 & 0 & m^2_{G^- \delta^+} \\ 
    0 & m^2_{H^{-}H^{+}} & m^2_{H^- \delta^+} \\ 
    m^2_{G^- \delta^+} &  m^2_{H^- \delta^+} & m^2_{\delta^- \delta^+} \\ 
  \end{array} \right)  \left(\begin{array}{c}
    G^+  \\ 
    H^+  \\ 
    \delta^+ \\ 
  \end{array} \right)\,, \label{eq:vm2}
 \ed
where the elements of mass matrix are found by
 \be
 m^2_{G^- \delta^+} &\approx & 0\,, \non \\
 m^2_{H^- H^+} &\equiv& m^2_{H^\pm} = \frac{m^2_{\pm}}{\sin\beta \cos\beta}\,,~m^2_{\pm} = m^2_{12}-\frac{\lam_4 + \lam_5}{2}v_1 v_2\,,\non \\
m^2_{H^- \delta^+} &=& \frac{v}{2\sin\beta \cos\beta} \left[\mu_1 \cos^4\beta -\mu_2 \sin^4\beta + (\mu_1 -\mu_2 ) \sin^2\beta \cos^2\beta \right] \,, \non \\
m^2_{\delta^- \delta^+} &\equiv& m^2_{\delta^\pm} = m^2_\Delta +\frac{v^2_1}{4} (2\lam_6 + \lam_7 + \lam_8) + \frac{v^2_2}{4} (2\bar\lam_6 + \bar\lam_7 + \bar\lam_8)\,.
\label{eq:MassMatrix}
 \ed
The null elements in Eq.~(\ref{eq:vm2}) are arisen from the neglect of small $v_\Delta$ that has been used in Eq.~(\ref{eq:mini}) for minimal conditions. Since $m^2_{G^- \delta^+}$ is also proportional to $v_\Delta$, for self-consistency, the $v_\Delta$ terms should be dropped. As a result, we get $m^2_{G^- \delta^+}\approx  0$, i.e. $G^\pm$ are the Goldstone bosons and decouple with $H^\pm$ and $\delta^\pm$. With this approximation, we find that the $3\times 3$ mass square matrix in Eq.~(\ref{eq:vm2}) could be reduced to be a $2\times 2$ matrix. The physical charged Higgs states could be regarded as the combination of $H^\pm$ and $\delta^\pm$ and their mixture  could be parametrized by 
 \be
  \left( \begin{array}{c}
    H^\pm_1\\ 
    H^\pm_2 \\ 
  \end{array}\right) =   \left(\begin{array}{cc}
    \cos\theta_\pm & \sin\theta_\pm \\ 
    -\sin\theta_\pm & \cos\theta_\pm\\ 
  \end{array}\right) \left( \begin{array}{c}
    H^\pm \\ 
    \delta^{\pm} \\ 
  \end{array}\right)\,.\label{eq:ma}
 \ed 
The masses of charged Higgs particles and their mixing angle are derived as
  \be
 \left( m_{H_{1,2}^{\pm}}\right)^2 &=&  \frac{1}{2}\left(m^2_{\delta^\pm} + m^2_{H^\pm}\right) \mp \frac{1}{2} \left[ \left( m^2_{\delta^\pm} - m^2_{H^\pm}\right)^2 + 4 m^4_{H^-\delta^+}\right]^{1/2}\,, \non \\
 \tan2\theta_\pm &=& - \frac{2 m^2_{H^- \delta^+}}{m^2_{\delta^\pm} - m^2_{H^\pm}}\,.
 \label{eq:mass_mixing}
  \ed
Here $H^\pm_1$  is identified as the lighter charged Higgs. 

Besides the couplings of $\delta^{\pm\pm}$ that exist in CTTSM, the scalar potentials in Eq.~(\ref{eq:v2}) provide new couplings to $H^\pm$. The relevant interactions could be found as
 \be
 -{\cal L}_{\delta^{\pm\pm} (H^\mp, \delta^\mp)}&=& \frac{1}{2} (2 \mu_1 + 2\mu_2) \delta^{++} H^- H^- - \frac{1}{2} (\sqrt{2} \lam_{10} v_\Delta) \delta^{++} \delta^{-} \delta^{-} \non \\
 &+& \frac{v \sin2\beta}{4} [(\lam_7-\lam_8)-(\bar\lam_7 - \bar\lam_8)] \delta^{++} H^- \delta^- +h.c.\,, 
  \label{eq:interaction}
 \ed
where  the first and third terms in RHS do not exist in CTTSM. If the charged Higgs $H^{\pm}$ is much lighter than $\delta^{\pm\pm}$, we see that the new decay channel $\delta^{++}\to H^+ H^+$ will be opened. Unlike the Feynman rules for  the interactions of  $\delta^{++}\ell\ell$ and $\delta^{++}W^-W^-$, the  new interactions are not   suppressed by  $m_\nu/v_\Delta$ or $v_\Delta$. In other words,  the decay rate of $H^+ H^+$ mode is much larger than that of $\ell^+ \ell^+$ and $W^+ W^+$; therefore,  the current  limit on the mass of doubly charged Higgs may be relaxed. Furthermore, the new decay channel $\delta^{++}\to H^+ W^{+^{(*)}}$ now is also allowed through the mixing angle $\theta_\pm$.
%

Next we discuss the numerical analysis for $\delta^{++}$ decays. According to the earlier discussions,  the relevant free parameters are angle $\beta$,  $\lam_{6,7,8}$, $\bar\lam_{6,7,8}$, $\lam_{4,5}$, $\mu_{1,2}$, $v$, $v_\Delta$, $m^2_{12}$ and $m_\Delta$. For reducing the free parameters and simplifying the numerical analysis, we take $v\approx 2 m_W/g$ as an input and assume $m_\Delta \sim m_{\delta^{\pm\pm}} \sim m_{\delta^\pm} $.  The involved parameters that we use for presentation are set to be angle $\beta$, $m_{H^\pm}$, $m_\Delta$, $v_\Delta$ and $\mu_{1,2}$.  Since the parameters $\mu_{1,2}$ are the important effects in our model, we adopt two different schemes for numerical discussions: (I) $\mu_1 = \mu_2=\mu$ and (II) $\mu_1= - \mu_2=\mu$. We note that by Eqs.~(\ref{eq:MassMatrix}) and (\ref{eq:mass_mixing}), the mixing angle $\theta_{\pm}$ is not a free parameter but is determined. Due to the tiny neutrino masses, the value of $v_\Delta$ is much less than 1 GeV.

\begin{figure}[hptb] 
\includegraphics[width=80mm]{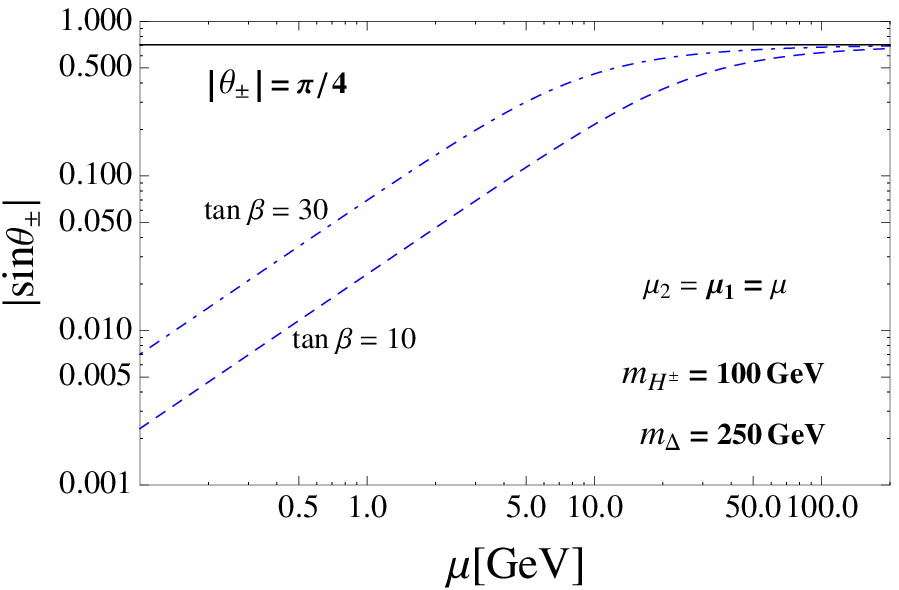}
\includegraphics[width=80mm]{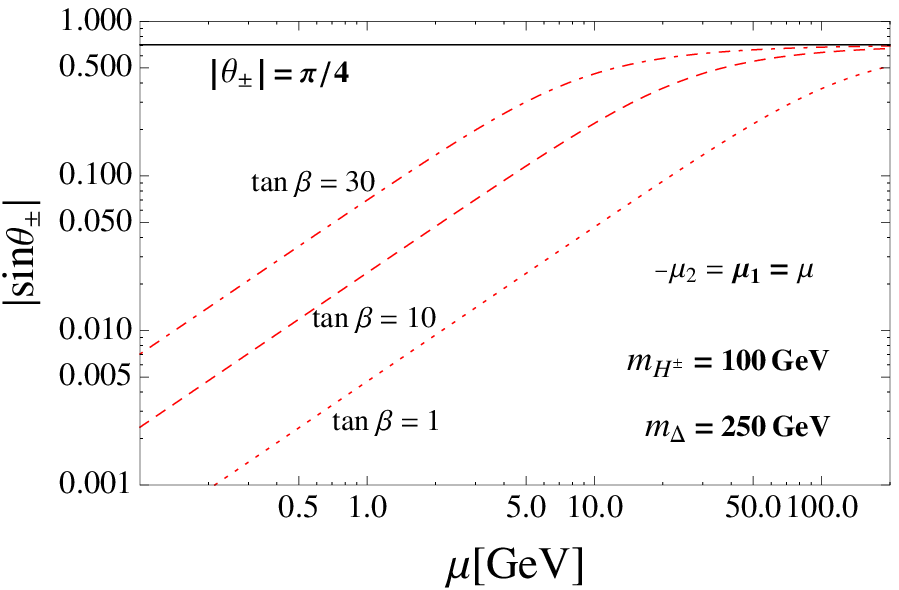} 
\caption{The mixing effect $|\sin\theta_\pm|$ of $H^\pm$ and $\delta^\pm$  as a function of $\mu$ with $m_{H^\pm} = 100$ GeV and $m_{\delta^\pm} =250$ GeV. The left panel is for scheme-I while the right panel is the scheme-II.  The dotted, dashed and dot-dashed line stands for  $\tan \beta =$ 1, 10 and 30, respectively.   \label{fig:mixing_angle}} 
\end{figure}

For understanding how the mixing angle $\theta_\pm$ depends on the free parameters, we plot $|\sin\theta_\pm|$ as a function of $\mu$ in Fig.~\ref{fig:mixing_angle}, where we have used $m_{H^\pm}=100$ GeV and $m_\Delta=250$ GeV, the left (right) panel denotes the scheme-I  (II), the dotted, dashed and dot-dashed line stands for $\tan\beta=1, 10, 30$, respectively, and the horizontal line corresponds to $|\theta_\pm| =\pi/4$. For scheme-I, due to $m^2_{H^-\delta^+}=0$ at $\tan\beta=1$, the mixing angle vanishes; therefore we only have two curves in the left panel.  By the plots, we see that when the value of $\mu$ is taken toward to $O(100)$ GeV, the mixing effect is approaching to  maximum. The value of $\mu$ cannot be arbitrarily large, otherwise the mass square of the lighter charged Higgs $H^\pm_1$ in Eq.~(\ref{eq:mass_mixing}) will become a negative.  
%
%

%
Now it is known that the magnitude of mixing effect of $H^\pm$ and $\delta^\pm$ strongly depends on the values of  $\mu_{1,2}$. We believe that  the interactions arisen from $\mu_j H^T_j i \Delta^\dagger H_j$ (j =1, 2) could lead to a new decay pattern for doubly charged Higgs. For more clarity,  we present the couplings of  $\delta^{\pm \pm}$ to the physical states $H^{\pm}_{1,2}$ and $W^{\pm}$  in 
Table~\ref{tab:interactions}. By the Table, we see that the involved free parameter for the vertex  $\delta^{\pm\pm}$-$H^\mp_1$-$W^\mp$ is only  the angle $\theta_\pm$.  Although the coupling for the vertex $\delta^{\pm\pm}$-$H^\mp_1$-$H^\mp_1$ could be comparable with that for  $\delta^{\pm\pm}$-$H^\mp_1$-$W^\mp$, due to phase space suppression,  the decay rate for $H^\mp_1 H^\mp_1$ mode usually will be smaller than that for $H^\mp_1 W^\mp$ mode, except the case with $\tan\beta=1$ and the case constrained by kinematic requirement.  
%
\begin{table}[hpbt]
\begin{center}
\begin{tabular}{cc|cc} \hline \hline
Vertex & Coupling & Vertex & Coupling \\ \hline \hline
 $\delta^{\pm \pm} H_2^\mp W_\mu^\mp $ & $-ig \cos \theta_\pm (p_{\delta^{\pm \pm}} - p_{H_2^\mp})_\mu$ 
 & $\delta^{\pm \pm} H_1^\mp W_\mu^\mp $ & $-ig \sin \theta_\pm (p_{\delta^{\pm \pm}} - p_{H_1^\mp})_\mu$ \\ \hline
 $\delta^{\pm \pm} H_{1(2)}^{\mp} H_{1(2)}^{\mp}$ & $2 (\mu_1+\mu_2) \cos^2 \theta_\pm (\sin^2 \theta_\pm)$ 
 & $\delta^{\pm \pm} H_1^{\mp} H_2^{\mp}$ & $2 (\mu_1+\mu_2) \cos \theta_+ \sin \theta_+$  \\ \hline \hline
\end{tabular}
\caption{The couplings of $\delta^{\pm\pm}$ to $H^{\pm}_{1,2}$ and $W^\pm$. 
 \label{tab:interactions}}
\end{center}
\end{table}
%
Applying these interactions,  the partial decay rates for $\delta^{\pm \pm} \to H^\pm_{1(2)} X$ ( $X=H^\pm_{1(2)}, W^\pm)$  could be formulated by 
\begin{align}
\Gamma (\delta^{\pm \pm }\to H^\pm_{1(2)} W^\pm) &= \frac{g^2 m_{\delta^{++}}^3}{16 \pi m_W^2} \sin^2\theta_\pm (\cos^2 \theta_\pm) 
\left[ \lambda \left( \frac{m_W^2}{m_{\delta^{++}}^2}, \frac{m_{H_{1(2)}^+}^2}{m_{\delta^{++}}^2} \right) \right]^{\frac{3}{2}}, \\
\Gamma(\delta^{\pm \pm }\to H^\pm_{1(2)} W^{\pm *} ) &= \frac{9 g^4 m_{\delta^{++}}^2 }{128 \pi^3} \sin^2\theta_\pm (\cos^2 \theta_\pm )
G \left( \frac{m_W^2}{m_{\delta^{++}}^2}, \frac{m_{H_{1(2)}^+}^2}{m_{\delta^{++}}^2} \right), \\
\Gamma(\delta^{\pm \pm} \to H_{1(2)}^\pm H_{1(2)}^\pm  ) &= \frac{(\mu_1+\mu_2)^2 }{4 \pi m_{\delta^{++}}^2 } \cos^4 \theta_\pm (\sin^4 \theta_\pm) \sqrt{m_{\delta^{++}}^2 - 4m^2_{H_{1(2)}^+}}, \\
\Gamma(\delta^{\pm \pm} \to H_{1}^\pm H_{2}^\pm  ) &= \frac{(\mu_1+\mu_2)^2 }{4 \pi m_{\delta^{++}}^2 } \sin^2 \theta_\pm \cos^2 \theta_\pm  
\sqrt{\lambda \left( \frac{m_{H_1^+}^2}{m_{\delta^{++}}^2}, \frac{m_{H_{2}^+}^2}{m_{\delta^{++}}^2} \right)},
\end{align}
where $W^{\pm *}$ expresses the off-shell W boson and
the functions $\lam(x,y)$ and $G(x,y)$, which are respectively associated with momenta of final particles and three-body phase space integration,  are found as~\cite{Aoki:2011pz}
\begin{align}
\lambda(x,y) =& 1+ x^2 +y^2 -2 xy -2x -2y \nonumber \\
G(x,y) =& \frac{1}{12 y} \biggl[ 2(x-1)^3 - 9(x^2-x)y + 6(x-1)y^2 \nonumber \\
& +6(1+x-y)y \sqrt{-\lambda(x,y)} \left( \arctan \left[ \frac{-1+x-y}{\sqrt{-\lambda(x,y)}} + \frac{-1+x+y}{\sqrt{-\lambda(x,y)}} \right] \right) \nonumber \\
& -3 [1+(x-y)^2-2y]y \log x  \biggr].
\end{align}
Since the doubly charged Higgs boson does not mix with other scalar bosons, the formulae for $\delta^{\pm\pm} \to \ell^\pm \ell^\pm$ and $\delta^{\pm\pm} \to W^\pm W^\pm$ decays are the same as those in CTTSM. Their explicit expressions could be found from Refs.~\cite{Aoki:2011pz, Perez:2008ha}.

Since there still involve  four new free parameters in our assumption, in order to  illustrate the characters of $\delta^{\pm\pm}$ in this model, 
we adopt several benchmark points (BPs) for the numerical analysis and they are given in Table~\ref{BPs1} (\ref{BPs2})
for scheme-I (II). In the Tables, we regard the values of $m_{\Delta, H^\pm}$, $\mu$ and $\tan\beta$  as inputs, then  $m_{H^\pm_{1,2}}$ and $\sin\theta_\pm$ are determined accordingly. 
%
\begin{table}[hpbt]
\begin{center}
\begin{tabular}{l||cccc||ccc} \hline \hline
 & $m_{\Delta}$ & $m_{H^\pm} $ & $\tan \beta$  & $\mu$  & $m_{H_2^\pm}$ & $m_{H_1^\pm}$ & $ |\sin \theta_+ |$ \\ \hline \hline
BP1 & 250 GeV & 100 GeV & 1 & 100 GeV   & 250 GeV & 100 GeV & 0  \\ 
BP2 & 500 GeV & 400 GeV & 10 & 100 GeV  & 579 GeV & 274 GeV & 0.57  \\
BP3 &  500 GeV & 400 GeV & 30 & 50 GeV  & 628 GeV & 124 GeV & 0.62 \\
BP4 &  120 GeV & 80 GeV & 10 & 5 GeV  & 133 GeV & 56 GeV & 0.47 \\
\hline \hline
\end{tabular}
\caption{Selected benchmark points in scheme-I.
 \label{BPs1}}
\end{center}
\end{table}
%
%
\begin{table}[hpbt]
\begin{center}
\begin{tabular}{l||cccc||ccc} \hline \hline
 & $m_{\Delta}$ & $m_{H^\pm} $ & $\tan \beta$  & $\mu$  & $m_{H_2^\pm}$ & $m_{H_1^\pm}$ & $ |\sin \theta_+ |$ \\ \hline \hline
BP5 & 500 GeV & 250 GeV & 1 & 100 GeV   & 503 GeV & 243 GeV & 0.13  \\ 
BP6 & 150 GeV & 100 GeV & 1 & 40 GeV  & 167 GeV & 68 GeV & 0.48  \\
\hline \hline
\end{tabular}
\caption{Selected benchmark points in scheme-II.
 \label{BPs2}}
\end{center}
\end{table}
%
%

\begin{figure}[hptb] 
\includegraphics[width=80mm]{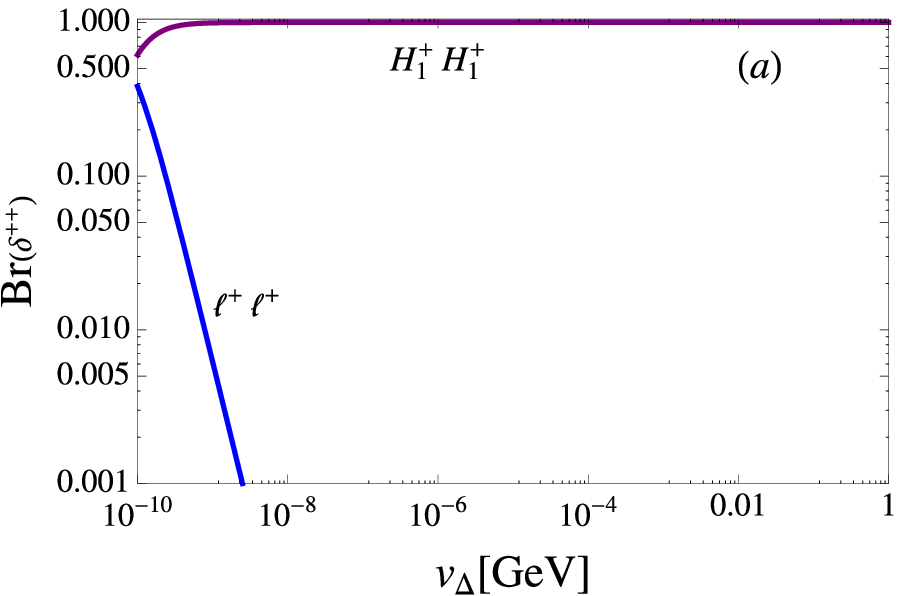}
\includegraphics[width=80mm]{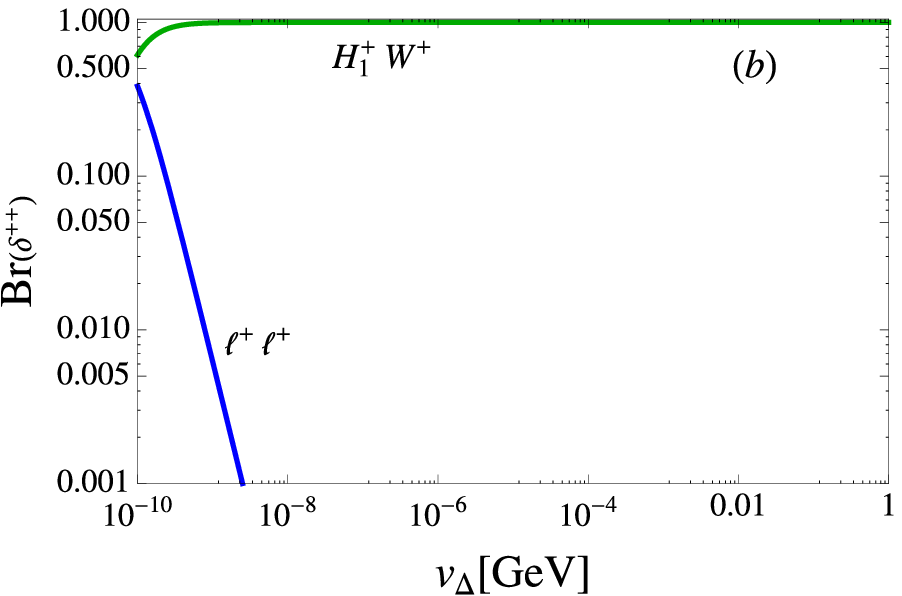} 
\includegraphics[width=80mm]{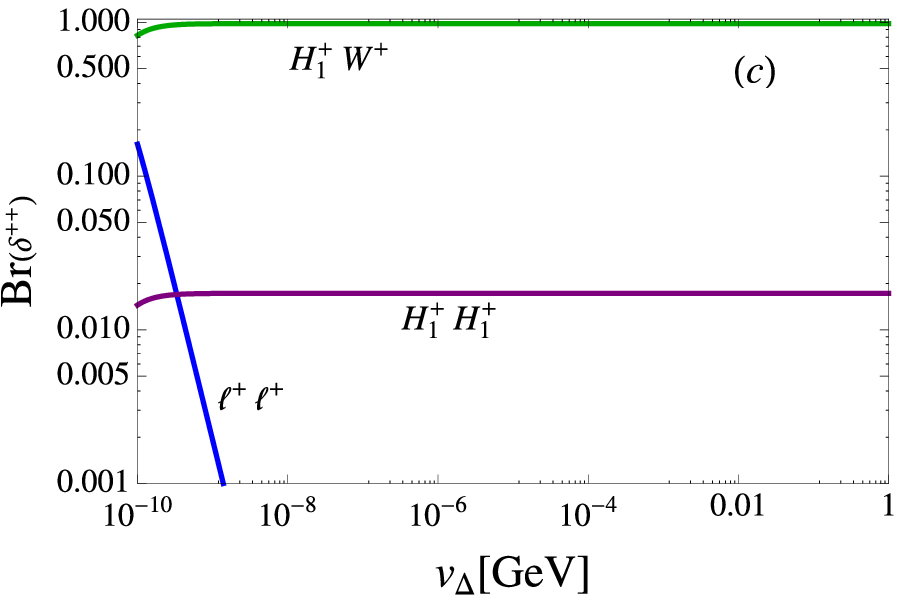} 
\includegraphics[width=80mm]{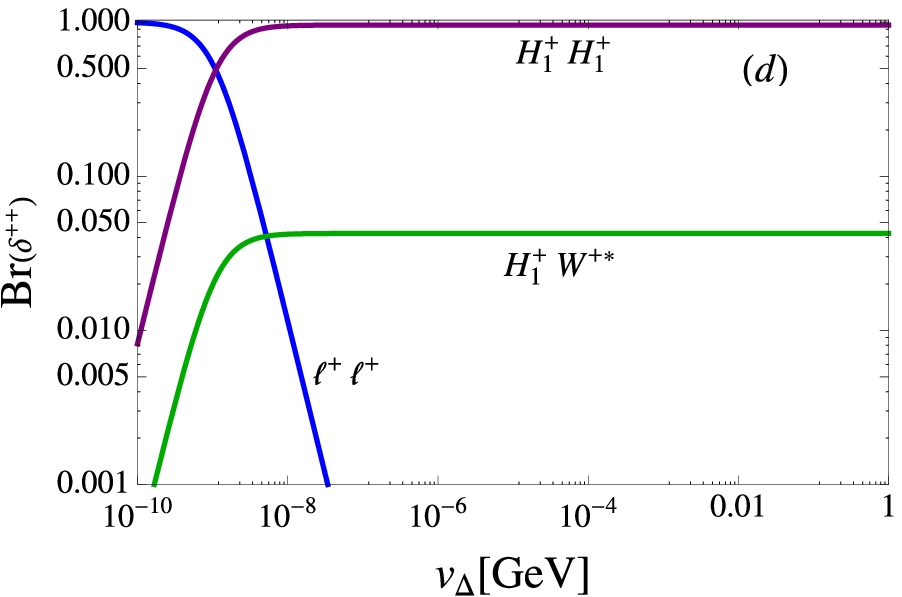}
\includegraphics[width=80mm]{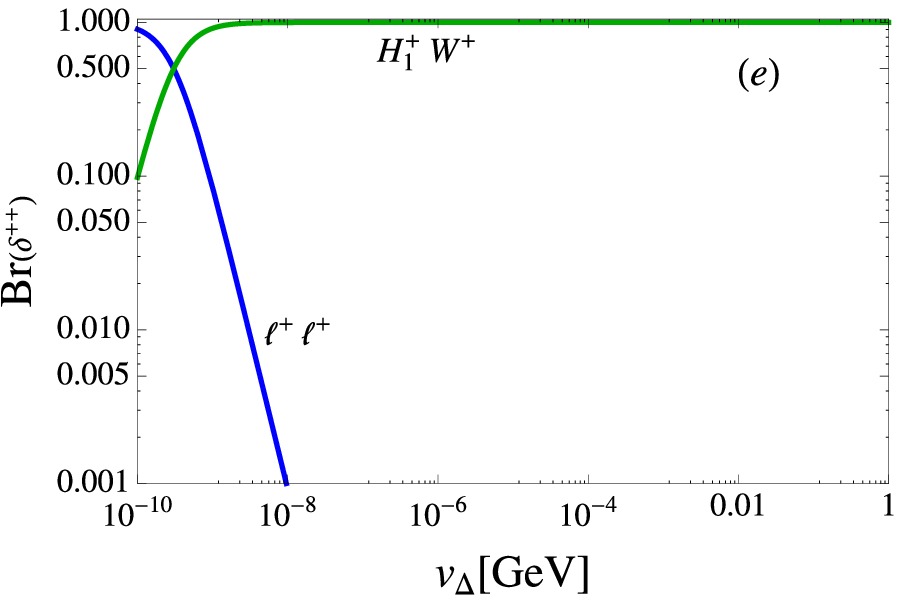} 
\includegraphics[width=80mm]{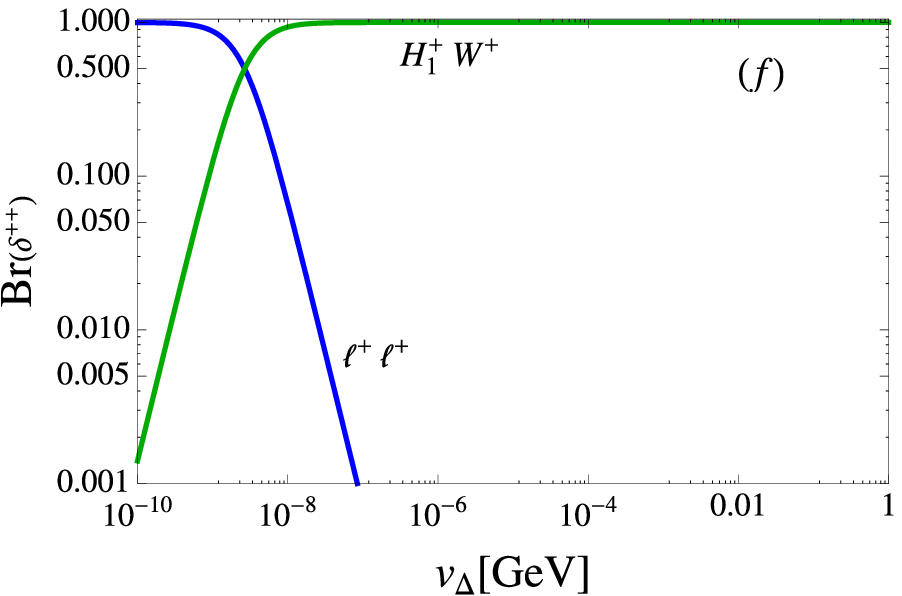}  
\caption{BRs for $\delta^{++}$ decays. (a)-(f) plot  respectively stands for BP1-6, defined in Tables~\ref{BPs1} and \ref{BPs2}.     \label{fig:Branching}} 
\end{figure}

In the following, we describe the characteristic  of each BP and display the associated results in Fig.~\ref{fig:Branching}. In BP1, we consider the case for $m_{\delta^{++}} > 2 m_{H_1^+}$ and set $\tan\beta=1$. Due to  $\theta_\pm =0$,  the decay $\delta^{++}\to H^+_1 W^+$ is suppressed. For comparison, we show the branching ratios (BRs)  for the decays $\delta^{++}\to (\ell^+ \ell^+,  H^+_1 H^+_1)$ in Fig.~\ref{fig:Branching}(a). In this paper, we use the normal ordering for neutrino masses to estimate  the decay rate of $\delta^{++}\to \ell^+ \ell^+$. By the plot, it is clear that the new open channel always dominates in the displayed region of $v_\Delta$. We note that in any circumstance, comparing with the new decay channel, $WW$ mode is very small and negligible. Hereafter, we will not mention the results of $WW$ mode. In BP2 and BP3, we select a heavier $m_{\Delta}$ and $\Delta m=m_{\Delta}-m_{H^\pm}=100$ GeV. From Table~\ref{BPs1}, we find that if we use $\mu\sim O(\Delta m)$, the mixing effect is $O(1)$ and the mass splitting between $m_{H^\pm_1}$ and $m_{H^\pm_2}$ is significant. Additionally, with larger value of $\tan\beta$, we see that the mixing angle and mass splitting are enlarged. We plot the  BRs of $\delta^{++}$ decays for BP2 and BP3  in Figs.~\ref{fig:Branching}(b) and (c). Since $m_{\delta^{++}} < 2 m_{H^+_1}$ in BP2, only $\ell^+ \ell^+$ and $H^+_1 W^+$ modes in Fig.~\ref{fig:Branching}(b) are allowed.  By the Figs.~\ref{fig:Branching}(b) and (c), we confirm the previous inference for $BR(\delta^{++}\to H^+_1 H^+_1)< BR(\delta^{++}\to H^+_1 W^+)$. In BP4, we use a lower mass for $m_{\delta^{++}}=120$ GeV and $m_{H^\pm}=80$ GeV. In this case, we find that the allowed value of $\mu$ cannot be over 7.7 GeV, otherwise $m^2_{H^+_1}$ will be negative.  Due to the kinematic requirement,  either $W^+$ and $H^+_1$ in $H^+_1 W^+$ mode should be off-shell. Since  the couplings of $H^\pm$ to quarks and leptons are related to the masses of fermions, for lighter charged Higgs decays, the decay rate for $\delta^{++}\to H^{+^*}_1(\to f_1 f_2) W^+$ is suppressed by the masses of lighter fermions. Therefore, we present the BRs for $\delta^{++}\to ( \ell^+ \ell^+, H^+_1 H^+_1, H^+_1 W^{+^*})$  in Fig.~\ref{fig:Branching}(d). Due to the phase space, we see  $BR(\delta^{++}\to H^+_1 H^+_1)> BR(\delta^{++} \to H^+_1 W^{+^*})$ in this case. Moreover, 
we also find that the decay $\delta^{++}\to \ell^+ \ell^+$ could become dominant when $v_\Delta$ is of order of $10^{-9}$.

For scheme-II,  the selected values of parameters are categorized in BP5 and BP6. Since the coupling of $\delta^{++} H^-_1 H^-_1$ vanishes in this scheme, the decay $\delta^{++} \to H^+_1 H^+_1$ is suppressed. Additionally,  the results with $\tan\beta=10$ and $30$ for $\delta^{++} \to H^+_1 W^+$ are similar to those in the scheme-I; therefore, we will not repeatedly  discuss the cases but focus on the case with $\tan\beta=1$. Hence, we present the BRs for $\delta^{++}\to (\ell^+ \ell^+, H^+_1 W^+)$ in Figs.~\ref{fig:Branching} (e) and (f),  where both BP5 and BP6 have similar behavior but the turning point of leading decay mode occurs at different value of $v_\Delta$. 
%
%

 It is known that the neutrinos get their masses at tree level in  type-II seesaw model. However, the neutrino masses could be also generated by loop corrections, e.g. the  two-loop effects which are similar to the Zee model \cite{Zee:1985id}. The loop corrections in CTTSM are actually negligible due to $\mu_i \sim v_\Delta \ll v$. Since we claim that  the $\mu_i$ could be as large as the VEV $v$, here it is worthy to discuss the loop effects in our model. 
%
By order of magnitude estimate, the two-loop effects are roughly expressed by
 \be
  m^\nu_{\ell'\ell} \sim \frac{1}{(4 \pi^2)^2 }\frac{m^2_{\ell'} m^2_\ell}{v^2 m^2_\Delta} ({\bf h})_{\ell' \ell} \mu_1 I_2 \,,
  \ed
  where we only consider the contribution of $\mu_1$ term, $1/(4\pi^2)^2$ denotes the two-loop effect, $m^2_{\ell'} m^2_{\ell}/v^2$ is from the vertex of $\bar L H_1 \ell_R$ in Eq.~(\ref{eq:lang_y}) and  the mass insertion in charged lepton propagators, 
${\bf (h)_{\ell' \ell}}$ is the Yukawa coupling of Higgs triplet in Eq.~(\ref{eq:lang_y}), and $I_2$ stands for the loop integration. 
Using $({\bf h})_{\ell' \ell} \sim  m_\nu/v_\Delta$, numerically we have $m^\nu_{\ell' \ell}\sim 10^{-12} (\mu_1/v_\Delta) (m^2_{\ell'} m^2_{\ell}/m^4_\tau)m_\nu I_2$.
Thus, by choosing proper value of $v_\Delta$, the radiative corrections to neutrino masses with $\mu_1 \sim O(v)$  could be still much smaller than the contributions from the tree level. 
%

Although our analysis focuses on the situation for which Higgs triplet is heavier than Higgs doublets, the reverse case should be also interesting and worth further studying. In the case of reversed mass ordering, the heavier doublet Higgs bosons can decay into doubly charged Higgs through 
charged Higgs decay $H^+ \to W^- \delta^{++}$ or through the cascade decay of neutral Higgs $H^0 \to H^+ W^- \to \delta^{++} W^- W^-$.   
Like SM Higgs, the heavier neutral Higgs would be produced by gluon fusion and have a sizeable production cross section at LHC.
%
Thus, it is interesting to search for the signal of the process $pp \to H^0 \to H^+ W^- \to \delta^{++} W^- W^-$ that  represents  the specific signature of the model.
Further studies of the collider signals are left as our future work. 

Finally, we give a remark on the couplings of triplet particles to quarks. As known that $\delta^{\pm}$ belongs to the $SU(2)$ triplet and cannot couple to quarks directly. However, the interactions of $\delta^{\pm}$ with quarks are built in our model through the mixing of $\delta^{\pm}$ and $H^\pm$, which is arisen from the $\mu_i$ terms of scalar potential.  Consequently, we open not only a new channel for the search of $\delta^{\pm\pm}$, but also a new way to look for it. For instance, if  $m_{\delta^{++}} \sim 250$ GeV and $m_{H^+_1}\sim180$ GeV, the signal for the existence of $\delta^{++}$ could be  read via the decay $\delta^{++}\to  H^{+^{(*)}}_1  W^+ \to t \bar b W^+ \to b \bar b W^+ W^+$, i.e. $2b$-jet+$W^+W^+$  in the final state,
%
%
%
%
where  the signal of $\delta^{++}$ becomes completely different from the CTTSM.

%

In summary, we have studied the new interactions in two-Higgs-doublet type-II seesaw model.  We find that the small VEV of Higgs triplet could be satisfied by accommodating the free parameters in the new scalar potential, i.e. $\mu_{1,2,3}$, $m_\Delta$, $v_{1,2}$, etc.,  where these massive parameters could be the same order of magnitude. By neglecting the contributions of $v_\Delta$, the charged Higgs mixing could be described by one mixing angle $\theta_\pm$. The mixing angle is dictated by the parameters $\mu_{1, 2}$ and $\tan\beta$.  We have demonstrated that by taking proper values of $\mu_{1,2}$ and $\tan\beta$,  the new decay channels  $\delta^{++}\to ( H^+_1 W^+, H^+_1 H^+_1)$  are dominant in $\delta^{++}$ decays, except at very tiny $v_\Delta$. Since the decay pattern of $\delta^{\pm\pm}$ is different from that in CTTSM, the search for $\delta^{\pm\pm}$ and the limit on its mass should be further studied at the colliders. It will be interesting to see the new phenomena  in the model at LHC.   \\

\noindent{\bf Acknowledgments}

 This work is supported by the Ministry of Science and Technology of 
R.O.C. under Grant \#: NSC-100-2112-M-006-014-MY3 (CHC) and NSC-102-2811-M-006-035 (TN). We also thank the National Center for Theoretical Sciences (NCTS) for supporting the useful facilities. 


\section*{APPENDIX: Mass matrices for CP-odd and CP-even Higgs bosons}

Using the potentials in Eq.~(\ref{eq:v2}) and the basis of Higgs doublets in Eq.~(\ref{eq:hH}), the mass matrix for the CP-odd components $G^0$, $A^0$ and $\eta^0$ is written by
\begin{equation}
\frac{1}{2} \begin{pmatrix} G^0 \\ A^0 \\ \eta^0 \end{pmatrix}^T
\begin{pmatrix} 
0 & 0 & m_{G^0 \eta^0}^2 \\ 
0 & m_{A^0 A^0}^2 & m_{A^0 \eta^0}^2 \\ 
m_{G^0 \eta^0}^2 & m_{A^0 \eta^0}^2 & m_{\eta^0 \eta^0}^2 
\end{pmatrix}
\begin{pmatrix} G^0 \\ A^0 \\ \eta^0 \end{pmatrix},
\label{M_CP_odd}
\end{equation}
where the elements of mass matrix are obtained as 
\begin{align}
\label{M_CP_odd_elements}
m_{G^0 \eta^0}^2 & \simeq 0  \nonumber \\
m_{A^0 A^0}^2 & \equiv m^2_{A^0}= \frac{m_{12}^2- \lambda_5 v_1 v_2}{\cos \beta \sin \beta} \nonumber \\
m_{A^0 \eta^0}^2 & = \frac{ v}{\sqrt{2} \cos \beta \sin \beta} [\mu_1 \cos^4 \beta - \mu_2 \sin^4 \beta + (\mu_1-\mu_2)\cos^2 \beta \sin^2 \beta ] \nonumber \\
m_{\eta^0 \eta^0}^2 & \equiv m^2_{\eta^0} =m_\Delta^2 + \frac{v_1^2}{2}(\lambda_6+\lambda_7)+ \frac{v_2^2}{2}(\bar{\lambda}_6+\bar{\lambda}_7).
\end{align}
The null elements in Eq.~(\ref{M_CP_odd_elements}) are arisen from the neglect of small $v_\Delta$ as in the charged Higgs case.
By using Eq.~(\ref{eq:mu3}), we get $m^2_{G^0 \eta^0} \propto v_\Delta$. Like the discussion on $m^2_{G^- \delta^+}$, for self-consistency, we should drop the $v_\Delta$  effect and take $m^2_{G^0 \eta^0} \approx 0$.
Thus, the mass matrix could be  reduced to a 2$\times$2 matrix. 
Consequently, the physical states of CP-odd Higgses could be parametrized by one mixing angle, defined by
\be
  \left( \begin{array}{c}
    A^0_1\\ 
    A^0_2 \\ 
  \end{array}\right) =   \left(\begin{array}{cc}
    \cos\theta_A & \sin\theta_A  \\ 
    -\sin\theta_A & \cos\theta_A \\ 
  \end{array}\right) \left( \begin{array}{c}
    A^0 \\ 
    \eta^{0} \\ 
  \end{array}\right)\,.\label{eq:ma}
 \ed 
 The masses of CP-odd Higgs particles and the mixing angle are derived as 
 \begin{align}
(m_{A_{1,2}^0})^2 &= \frac{1}{2}(m_{A^0 A^0}^2 + m_{\eta^0 \eta^0}^2) \mp \frac{1}{2} \sqrt{(m_{A^0 A^0}^2-m_{\eta^0 \eta^0}^2)^2 + 4 (m_{A^0 \eta^0})^4 } \,, \non \\
\tan 2 \theta_A & = \frac{2 m_{A^0 \eta^0}^2}{m_{A^0 A^0}^2-m_{\eta^0 \eta^0}^2}\,,
\end{align}
where $A_1^0$ is identified as the lighter CP-odd Higgs.

For CP-even Higgs bosons, first we transform the $h^0$ and $H^0$ states to $h$ and $H$ states by
 \begin{equation}
\begin{pmatrix} h^0 \\ H^0 \end{pmatrix} 
= 
\begin{pmatrix}  \cos \alpha &  \sin \alpha \\ -\sin \alpha & \cos \alpha \end{pmatrix}
\begin{pmatrix} h \\ H \end{pmatrix}\,,
\end{equation}
where $h$ and $H$ usually are the physical mass eigenstates in THDM and  $\alpha$ is the mixing angle. 
With Eq.~(\ref{eq:hH}), 
 we write $\rho_{1, 2}$ in terms of $h$ and  $H$ as
\begin{equation}
\begin{pmatrix} \rho_1 \\ \rho_2 \end{pmatrix} 
= 
\begin{pmatrix} \cos (\alpha-\beta) & \sin (\alpha -\beta) \\  -\sin (\alpha-\beta) & \cos (\alpha-\beta) \end{pmatrix}
\begin{pmatrix} h \\ H \end{pmatrix}
.
\end{equation}
In this basis, the mass matrix becomes 
\begin{equation}
\frac{1}{2} \begin{pmatrix} h \\ H \\ \delta^0 \end{pmatrix}^T
\begin{pmatrix} 
m^2_{hh} & 0 & m_{h \delta^0}^2 \\ 
0 & m_{H H}^2 & m_{H \delta^0}^2 \\ 
m_{h \delta^0}^2 & m_{H \delta^0}^2 & m_{\delta^0 \delta^0}^2 
\end{pmatrix}
\begin{pmatrix} h \\ H \\ \delta^0 \end{pmatrix}.
\label{M_CP_even_2}
\end{equation}
The elements of the mass matrix and $\tan 2 \alpha$ are given by 
\begin{align}
m_{HH,hh}^{2} &=  \frac{1 }{2} \left[ m_{12}^2 \left(\tan \beta + \cot \beta \right) +2(\lambda_1 \cos^2 \beta + \lambda_2 \sin^2 \beta) v^2 \right]  \nonumber \\
& \pm \frac{1}{2} \sqrt{ \left[ m_{12}^2 \left(\tan \beta - \cot \beta \right) +2(\lambda_1 \cos^2 \beta - \lambda_2 \sin^2 \beta) v^2 \right]^2 + 4 (m_{12}^2 - \lambda_{345} v^2 \sin \beta \cos \beta)^2  }  \,,
\nonumber \\
m_{H \delta^0}^2 & = \frac{v}{\sqrt{2}} (\mu_1 \cot \beta - \mu_2 \tan \beta) \sin (\alpha - \beta)\,,  \nonumber \\
m_{h \delta^0}^2 & = \frac{v}{\sqrt{2}} (\mu_1 \cot \beta - \mu_2 \tan \beta) \cos (\alpha - \beta) \,, \nonumber \\
\tan 2 \alpha & = \frac{2 (- m_{12}^2 + \lambda_{345} v^2 \sin \beta \cos \beta) }{m_{12}^2 \left(\tan \beta - \cot \beta \right) +2(\lambda_1 \cos^2 \beta - \lambda_2 \sin^2 \beta)v^2}\,,
\end{align}
where we have used Eq.~(\ref{eq:mu3}) and $\lambda_{345} = \lambda_3 + \lambda_4 + \lambda_5$.
Since $m^2_{h\delta^0}$ is not suppressed by $v_\Delta$, the  3$\times$3 mass matrix in general cannot be further reduced. 
However,  for  the case with $\sin (\alpha-\beta) \sim - 1$ where $h$ is the SM-like Higgs particle,  due to  $m_{h \delta^0}^2 \sim 0$,
 the mass matrix then could be reduced to a 2$\times$2 mass matrix.  In sum, due to the $\mu_i$ terms, large mixing effects between triplet and doublet particles occur in our model.  


\end{document}